# LARGE - SCALE HOMOGENEITY OR PRINCIPLE HIERARCHY OF THE UNIVERSE?

**Chechelnitsky A.M.** Laboratory of Theoretical Physics,
Joint Institute for Nuclear Research,
141980 Dubna, Moscow Region, Russia
E'mail: ach@thsun1.jinr.ru

## ABSTRACT

Invariablly justified representations of the Wave Universe Concept - WU Concept (See monography - Chechelnitsky A.M. Extremum, Stability and Resonance in Astrodynamics..., etc. and other publications) indicate a principle incorrectness of expectations of Standard (Model) cosmology about homogeneity and isotropy of the Universe.

It also is connected with observational data about apparent hierarchy of giant astronomical systems (stellar systems, galaxies, clusters of galaxies, superclusters of galaxies, etc.), their megawave structure, quantization "in the Large", non-homogeneity of microwave background Space radiation, adequately interpreted (in frameworks WU Concept) effects redshifts quantization of quasars, etc.

The principle absence of a Limit of Hierarchy of Matter Levels asserts: "The Staircase of a Matter" - is endless.

For orientation of the explorers, working with the observational data, in frameworks of WU Concept the concrete characteristics of following (behind superclusters of galaxies) potentially possible extremely large astronomical systems are calculated with using the Fundamental parameter of Hierarchy – Chechelnitsky Number $\chi = 3.66(6)$.

The astronomical systems, belonging to the nearest hierarchy Levels of Solar-Like systems, are characterized by external radiuses $[a^{(k)} = \chi^k a^{(0)}, a^{(0)} = 39.373 \text{ AU}]$
$$a^{(20)} = 36.83, \; a^{(21)} = 135, \; a^{(22)} = 495, \; a^{(23)} = 1815 \text{ Mpc}.$$

It is possible to expect that in the Universe also exist and should show itself in observations (the Solar-Like objects) – extremely large astronomical systems (ELAS), characterized by the external radiuses (of peripherals)
$$a^{(26)} = 89503, \; a^{(27)} = 328177, \; a^{(28)} = 1203318 \text{ Mpc}.$$

## COSMOLOGY - STANDARD MODEL:
### At Last There Will Set of Long-Awaited Homogeneity...

Almost whole century the cosmology of new time is in painful expectation: It demands from observational astrophysics to confirm a postulate, similar to dream, of the refined theory: The Universe - is *homogeneitic* and is *isotropic*.

Unfortunately, for the supporters of Standard model the triumph of the prescriptions of the prevailing theory permanently is sidetracked. It - lengthy and (would be desirable to believe) instructive history.

## TOWARDS TO (MEGA) WAVE UNIVERSE

### What - Beyond the Horizon of (Visible) Universe?

Today, being grounded on competitive representations of Standard Model of a Cosmology and of the Wave Universe Concept (WU Concept) it is interesting to attempt to answer on following (probably, anticipatory and impatient) problem:

What will meet tomorrow (in XXI century - in III millenary) grown-up, more technically equipped and, probably, conceptually more perfect astronomy, astrophysics, cosmology in the Universe - *Monotonic Desert of homogeneous "gas" (clusters, superclusters) galaxies*?

Or, "having tightened" for some Levels of Matter, it will find out new, more extended consolidated, close to stationary, *astronomical objects of the extremely large sizes and masses*?

**What - Behind a World of Superclusters of Galaxies?**

In other words, whether there will be a cosmology hereafter again opened Levels and Layers of a Matter, physically isolated astronomical systems superior the sizes of observed at present superclusters of galaxies?

Not sidetracking, we at once, here and now, attempt to answer this problem, being grounded on representations of the Wave Universe Concept (WU Concept).

So,

\# It is necessary to expect, that purposeful researches and the future successes of an observational astrophysics really will lead to detection of astronomical objects more high rank, than superclusters of galaxies.

\#Potentially possible, to the greatest degree probable characteristics of these extremely large astronomical objects can be indicated as outcome of the analysis in frameworks of WU Concept.

## THE WAVE UNIVERSE CONCEPT (WU CONCEPT).
## WAVE ASTRODYNAMICS

The *Wave Universe Concept (WU Concept)* and fundamental ideas of *Wave astrodynamics* [Chechelnitsky, 1990 -1999] are connected with representations that the large astronomical systems in the theoretical plan are not only multiparticle dynamic systems in sense Poincare-Birkgof, but are considered as essentially *Wave Dynamic Systems* (WDS), systems, being somewhat analogues of atom.

**The Fundamental Wave Equations.**
**Stability, Quantization of Megasystems**

The theoretical aspects of these problems (in particular, eigenproblem of the *Fundamental wave equations*) both appropriate astronomical and astrophysical questions are discussed in the monography [Chechelnitsky, 1980] and subsequent publications.

## SHELL STRUCTURE OF ASTRONOMICAL SYSTEMS

The any astronomical systems of the Universe considered as *wave dynamic systems* (WDS) have shell structure, in many respects similar with *shell structure* of Solar - planetary system [Chechelnitsky, 1980, 1983-1986].

The exceptions in this sense and numerous satellite systems of planets do not constitute, it is good verified by experience, observations and space experiments.

**Shell Hierarchy**

In that case, the astronomical system considered as WDS, is characterized by *hierarchy* enclosed each other spatially and structurally (radially) of the divided areas - $G^{[s]}$ *Shells* (s = ..., -2, -1, 0, 1, 2, 3...).

Inducing experience in research of *wave shell structure* of any astronomical systems are the results of an experimental research of Solar system - most in details and authentically known astronomical system.

In Solar - Planetary system some spatially divided shells - can be clearly identified, at least -

$G^{[0]}$ - Intra - Mercurian;
$G^{[1]}$ - taken by space of planets I (Earth) group;
$G^{[2]}$ - taken by space of planets II (Jupiter) group;
$G^{[3]}$ - Trans - Pluto etc.

**Sound Velocities Hierarchy.**
**Fundamental Parameter of Hierarchy**

The hierarchy of the $C_*^{[s]}$ *sound velocities* - phase velocities of the (multicomponent cosmic medium) cosmic plasma small perturbations (*megawaves*) [Chechelnitsky, 1984,1986] is closely connected with the hierarchy of $G^{[s]}$ Shells

$$C_*^{[s]} = (1/\chi^{s-1}) \cdot C_*^{[1]}, \quad (s = ..., -2, -1, 0, 1, 2, 3,...),$$

where $C_*^{[1]} = 154.3864$ км·s$^{-1}$ is the calculated value of sound velocity in the $G^{[1]}$ Shell, that was made valid by observation, and

$\chi = 11/3 = 3.66(6)$ – is the *Fundamental parameter of hierarchy* (*Chechelnitsky Number*) [Chechelnitsky, (1978) 1980 -1988].

**"Magic Number" (Chechelnitsky Number, FPH) $\chi$=3,66(6).**
**Role and Status of Fundamental Parameter of Hierarchy in Universe.**

Previous after primary publications [Chechelnitsky, 1980-1985] time and new investigations to the full extent *convince* the theory expectations, in particular, connected with the $G^{[s]}$ Shells hierarchy in each of such WDS, with the hierarchy of Levels of matter (and WDS) in Universe, with the exceptional role of the *introduced in the theory* $\chi$ FPH [Chechelnitsky, (1978) 1980-1986].

The very brief resume of some aspects of these investigations may be formulated in frame of following short suggestion.

**Proposition (Role and Status of $\chi$ FPH in Universe)** [Chechelnitsky, (1978) 1980-1986]

# The central parameter, which organizes and orders the dynamical and physical structure, geometry, hierarchy of Universe

* "*Wave Universe (WU) Staircase*" of matter Levels,
* *Internal* structure each of real systems - wave dynamic systems (WDS) at *any Levels* of matter, is (manifested oneself) $\chi$ - the *Fundamental Parameter Hierarchy (FPH)* - nondimensional number $\chi$ =3,66(6).

# It may be expected, that investigations, can show in the full scale, that $\chi$ - FPH, generally speaking, presents and appeares *everywhere* - in any case, - in an extremely wide circle of dynamical relations, which reflect the geometry, dynamical structure, hierarchy of real systems of Universe.

We aren't be able now and at once to appear all well-known to us relations and multiple links, in which oneself the [Chechelnitsky] $\chi$=3.66(6) *"Magic Number"* manifests.

We hope that all this stands (becomes) possible in due time and with new opening opportunities for the publications and communications.

# HIERARCHY STRUCTURE OF WAVE UNIVERSE.
# STEPS OF HIERARCHY. A STAIRCASE OF MATTER

**Hierarchy of Solar - Similar Systems.**

According to representations of the Wave Universe Concept (WU Concept) the Hierarchy of Solar - similar systems (Solar-like Systems - SL Systems) can be shown, first of all, by *Homology* - by Homologous series of *Main dynamic parameters - parameters of the Kepler*

$$K^{(k)} = \chi^k K^{(0)} \qquad k = \ldots, -2, -1, 0, 1, 2, \ldots$$

where

$K^{(0)} = K_\odot = 1.32712438 \cdot 10^{11}$ км$^3 \cdot$s$^{-2}$ - Main dynamic parameter - parameter of the Kepler - *Gravity parameter of the Solar System (Sun)*;

$\chi = 3.66(6)$ - Fundamental parameters of hierarchy (Chechelnitsky Number) [Chechelnitsky, (1978) 1980, 1980-1986];

$k = 1, 2, 3, \ldots$ - countable parameter.

The Sun and Solar system are most well investigated objects of a population of stars and planetary systems, in many *respects are typical*, steady enough, *well observed* representatives of a *Layer (and Levels) of Matter* - stars - one of the brightest component of a Staircase of Matter. It is necessary to expect, that been *transposed* in other Levels of Matter, the representatives of a Homologous series $K^{(k)}$ also will appear by reference, *quite typical*, steady enough, it is *good and widely observed* objects.

In other words, it is necessary to expect, that $K^{(k)}$ and at other Levels of a Matter $U^{(k)}$ also will appear *quite representative*, *widespread* objects, such as the Sun and Solar system.

These expectations can be affirmed by the observational data.

**The Basis for Selection of Solar-Like Systems.**

We are reverted to the analysis of Hierarchy of Solar - like systems, at least, by virtue of following circumstances:

# *Determinancy*.

Dynamic and physical properties of the Sun and the Solar systems are known with *extraordinary accuracy* (as contrasted to by set of diverse astronomical objects). It means, as all Homologous series of Solar-similar systems is representable quite definitely with *reasonable accuracy*.

# *Representativity*.

The Sun - as an star is the *quite typical representative* of a Layer of stars in a Staircase of a Matter. About it speaks, for example, a median *position* of the Sun on the Hertzsprung-Russel diagramm. It is necessary to expect, that other generated components of a Homologous series of Solar–like systems (for example, in a *Layer of galaxies*) also will be *quite representative* objects at the conforming Levels of a Matter.

**Hierarchy of External Sizes of Systems (Pluto-Like Orbits).**

As a subject for the analysis it is interesting to consider also Homologous series of Pluto-Like Orbits - PL Orbits, External PL Sizes, connected with Hierarchy of Solar - like systems.

If to accept as a *generating component (Eponym)* the semi-major axis of orbit of Pluto (P) in a Solar System (SS)

$$a^{(o)} = a_{SS,P} = 39.37364 \text{ AU} = 0.0001900889 \text{ pc},$$

than a *Homologous series of external (PL) sizes* of Solar-Like systems will look like this:

$$a^{(k)} = \chi^k a^{(o)} = \chi^k a_{SS,P} = \chi^k 39.37364 \text{ AU} = \chi^k 0.000190889 \text{ pc},$$

where $k = 1, 2, 3, \ldots$ - countable parameter,

$\chi = 3.66(6)$ - Fundamental parameter of hierarchy (Chechelnitsky Number).

It is necessary to expect, that, similarly to the semimajor axis (radius) $a^{(o)} = a_{SS,P}$ of Pluto orbit, reflecting the peripheral size of a Solar system, the radiuses $a^{(k)}$ of a Homologous series of external (PL) sizes will describe (maximal) peripheral - *external sizes* - radiuses of astronomical systems of corresponding $U^{(k)}$ Levels of Matter. Its are the most simplly and directly observed in an astrophysics values – *linear dimensional* characteristics of astronomical systems - clusters of stars, galaxies [see the Table].

**Extremely - Large Astronomical Systems. First Steps.**

Analysing a Homologous series of external (PL) sizes, it is possible at once directly to indicate potentially possible existence in the Universe of extremely - large astronomical systems (objects).

It is possible to expect, for example, that in the Universe exist and should show itself in observations (Solar-Like objects) - *extremely large astronomical systems (ELAS)*, characterized by the sizes (external) radiuses of peripherals

$$a^{(26)} = \chi^{26} a_{SS,P} = 89503 \text{ Mpc},\ a^{(27)} = 328177 \text{ Mpc},\ a^{(28)} = 1203318 \text{ Mpc}.$$

It - apparently, steady enough objects of new, nearest, potentially existing (ELAS) *Layer of Matter*. The distinguished nature of these pointed Levels of Matter in new (ELAS) Layer of Matter can be realized also from following simple heuristic consideration.

If to consider a Level of Matter - Solar system (Layer of stars) (external size) in any dynamic sense conforming to a Level of Matter - of our Galaxy (or of galaxy M31 Andromeda) (Layer of galaxies) (external size $a_G$), than for following higher (ELAS) Layer of Matter (the external size $a_{ELAS}$) is possible to record a following ratio (of similarity) of external sizes of astronomical systems

$$a_G / a_{SS,P} \sim a_{ELAS} / a_G$$

From here we have
$$a_{ELAS} \sim (a_G)^2 / a_{SS,P}$$

In more detail such conclusion follows from consideration, connected with definite *isomorphism* of Layers of Matter.

**Stability - Observability.**

The general reasons about a problem "Stability - Observability" can be found out in the monography (Chechelnitsky, 1980).

But the problem of stability, so, and actual observability of objects of high Levels of hierarchy represents the special, non trivial problem. It merits the special discussion.

**Isomorphism of Layers of Matter - Stars and Galaxies.**

From the point of view of WU Concept representations the hierarchical structure of themselves Layers of Matter in definite dynamic sense - is *look-alike*.

The definite *similarity - isomorphism* of Layers of a Matter - stars and galaxies can be observed if to compare, for example, Levels of Matter $U^{(0)}$ of a Layer of stars and $U^{(13)}$ (or $U^{(14)}$) of Layer of galaxies and further, accordingly, subsequent Levels of Matter ($U^{(1)}$ and $U^{(14)}$, $U^{(2)}$ and $U^{(15)}$, etc.).

In that case it appears, that the aggregates - *clusters of galaxies* (clusters, clusters of clusters, etc) correspond to aggregates - *clusters of stars* (clusters of stars, spherical clusters, etc). The capability also opens to present dynamic structure nowadays of unknowns at high Levels of a Layer of Matter - galaxies on the basis of comparison with dynamic stucture of known high Levels of Matter of a Layer of Matter - of *stars.*

**The Nearest Levels of Matter.**

But today, apparently, the problem of existence of grandiose astronomical systems located on the *nearest* steps of Hierarchy of the Universe is most actual. It is dictated by capabilities and technical limitations of a modern observational astrophysics.

If to consider, that at present clearly identifiable astronomical systems (the superclusters of galaxies) are characterized by external sizes - radiuses of the order

$$a \sim 30\ h^{-1}\ \text{Mpc}\quad \text{(Peebles, 1980 (1983))},$$
$$a \sim 35\ \text{Mpc}\quad \text{(LSS: Rudnicki, Zieba)},$$
$$a \sim 40(50)\ \text{Mpc (LSS: Kalinkov et al)}$$

that its, most likely, belong to (or are close to) a Level of Matter characterized by an external radius of a Solar–Like system $a^{(20)} = \chi^{20} a_{SS,P} = 36.83$ Mpc.

The astronomical systems, belonging to the nearest hierarchy Levels of Solar-Like systems, in that case are introduced by external radiuses

$$a^{(21)} = 135\ \text{Mpc},$$
$$a^{(22)} = 495\ \text{Mpc},$$
$$a^{(23)} = 1815\ \text{Mpc}$$

It is interesting to mark, that some explorers working with the observational data (LSS: Rudnicki, Zieba), indicate *distinguished nature* of the following sizes (radiuses)

$$a \sim\ 35,\quad 128,\quad 421\ \text{Mpc}.$$
(See also Einasto, 2000 with $a \sim 130$ Mpc).

Whether is contingency the close conformity of these data obtained from processing of the cataloques (Abell, Zwicky) of galaxies clusters with the analysis and expectations of the Wave Universe Concept?

**The Law of Generalized Dichotomy and External Sizes of Astronomical Systems.**

At statistical processing of the cataloques of galaxies clusters it find out a series of maxima in distribution of the characteristics of the sizes of galaxies aggregates (clusters). Thus at processing the *isolated nature* of these aggregates implicitly is meant.

Agrees [LSS: Rudnicki, Zieba], this series of maxima - preferential sizes looks like

$$a_{obs} = 35, 63, 64, 69, 72, 82, 93, 128, 142, 190, 206, 289 \text{ Mpc}.$$

In frameworks WU Concept the latent sense of such distribution can be realized. The preferential sizes of aggregates (clusters) of galaxies appear by connected with *the Law of generalized dichotomy*

$$a_p = a_o \cdot 2^{p/2}, \quad p = 0, 1, 2, 3, \ldots,$$

characteristic for the dominant sizes (and other parameters) of astronomical systems (Chechelnitsky, 1984, 1992, 1999). This Law is some generalization (in frameworks of WU Concept) of known Law a Titius-Bode for planetary orbits.

In this connection it is interesting to mark following conformity of the observational data with series of Generalized Dichotomy at ($a_o = 36$ Mpc $\sim a^{(20)} = 36.83$ Mpc)

$$a_p = 36 \cdot 2^{p/2} \text{ Mpc}, \quad p = 0, 1, 2, 3, \ldots,$$

| $a_p$ | 36 | 50.91 | 72 | 101.82 | 144 | 203.6 | 288 Mpc |
|---|---|---|---|---|---|---|---|
| $a_{obs}$ | 35 | | 72 | | 144 | 206 | 289 Mpc |

Besides it is necessary to mark individual ratio of dichotomy for the falling out observational data

$$a_{obs} \sim 64, 122 \text{ Mpc} \rightarrow 2 \times 64 \text{ Mpc} = 128 \text{ Mpc}$$
$$a_{obs} \sim 93, 190 \text{ Mpc} \rightarrow 2 \times 93 \text{ Mpc} = 186 \text{ Mpc} \sim 190 \text{ Mpc}$$

**Causing for Hope for Homogeneity?**

The researches of last time, generally speaking, have not changed the information on availability of heterogeneities - distinguished scales (maxima) of clustering of galaxies.

But the volume of the processed data considerably has increased, as well as quantity of a usedprocessing techniques. It gives the basis to some explorers [Kalinkov et al. 1998] to suppose, for example, following:

" It seams that there are *no structures* of superclasters of galaxies ".

It is understandable, that such conclusion lies quite in a channel of Standard model and owes, in next time, with gladness to be hailed by the representatives of Standard cosmology Mainstream.

Unfortunately, this brief conclusion not absolutely corresponds even to the contents of this large and valuable work. At desire in it is possible to see, for example, availability of *distinguished scales* (see Kalinkov t al., 1998, Fig. 1,2,3)

$$a_{obs} \sim 90 h^{-1} \text{ Mpc}, 325 h^{-1} \text{ Mpc} \quad \text{etc.}$$

At a Hubble constant H=65 км·s$^{-1}$·Mpc$^{-1}$ (h = 0.65) it gives distinguished scales,

$$a_{obs} \sim 138, 500 \text{ Mpc},$$

quite comparable with external radiuses of giant astronomical systems

$$a^{(21)} = 135 \text{ Mpc}, \quad a^{(22)} = 495 \text{ Mpc}$$

Generally speaking, it is follows with fear to approach to results of statistical processing of rather vast, but heterogeneous stuff. Occasionally, the desired signal (availability of maxima) is lost ("is washed") in a massif of such rich data set. Be it – *case history to a maxim* "best - enemy of good".

In any case, detail critical analysis of such works merits separate, special discussion.

**View Ad Infinitum:**
**About a Capability of Existence of Extreme - Large Astronomical Systems.**

If not to limit by consideration only of nearest Levels and Layers of Matter, the Wave Universe Concept gives a capability to investigate (for the present - theoretically) an hierarchical stucture of the Universe, in principle, at any Levels and Layers of Matter. All Hierarchy of the Wave Universe, all Staircase of Matter - potentially is opened for free and unbiased researches object.

This polygon for possible researches extends (even at present familiar - with verifiable modern experiments – physical world) many tens orders: downwards - deep into Matter in subatomic world and hill up - in world of extreme – large astronomical systems.

In this connection the fundamental (in definite sense - epistemological or metaphysical) statement of WU Concept is imply following:

**The Assertion (Hierarchy Ad Infinitum)**
There is No limit of Hierarchy (of Staircase of Matter).
This extreme brief statement, at least, does not limit by dogmas the horizon of a cosmology of the Future, does not bar creative, search tendencies of the explorers.

**DISCUSSION**

The Standard Model of a modern cosmology does not test the special doubts that the Universe "As a whole" ("All Universe" - in the issue) is homoheneitic and is isotropic. But the impartial and critical analysis demonstrates, that such reliance, mainly, reposes on formerly none-critically adopted by the fathers-founders (of modern cosmology) postulates, and later - it is already simple – on tradition of following to authorities, habit, mode.

The scandalous gap with observed properties of the actual Universe is eliminated by the ad hoc confirmation that nevertheless at a definite stage in time and space the Universe (why?) *ceases to be homogeneous* and therefore appear reference observed features an actual hierarchic world - the world of atoms, planetary, star galactic systems.

The Wave Universe Concept tumbles this model and puts it from a head on legs, first of all, - on the *apparent, fundamental observational basis*. The observed Hierarchy of Levels of Matter extends many tens orders and there are no visible causes, by virtue of which one this hierarchy should *interrupt* at any Level of Matter. You see then it will be special, *really physically distinguished* Level of Matter. Here again not only it is desirable, but it is necessary to result extremely severe, nontrivial, convincing physical arguments, by virtue of which one the so series *universal property of hierarchy of Universe* is for some reason upset.

Moreover, the Wave Universe Concept indicates the causes, circumstances, arguments, by virtue of which one the *phenomenon of Hierarchy* is an indispensable, natural consequent of some more fundamental laws of nature. Briefly speaking, the observed Hierarchy of the Universe is to straight lines a consequent of a *Wave (Megawave)* constitution of the actual Universe, consequent that it is grandiose composition of the enclosed *Wave dynamic systems* (WDS) at all Levels of Staircase of Matter.

Thus the immanent wave (megawave) properties spontaneously and directly are connected to properties of quantization (including, and quantization "in the Large"), commensurability (as inside WDS, and between them - at miscellaneous Levels of Matter) and, thus - with properties and laws of Hierarchy.

In order to prevent barren loss of time and efforts, cosmologists, which one with the tight attention prolong to expect approach of epoch of total homogeneity and isotropy of the Universe, it is necessary to advise - more often to recall astern wisdom: "If you very wait for the friend, do not accept knocking the heart for stamp of hoofs of his horse".

## HIERARCHY OF UNIVERSE. STAIRCASE OF MATTER.
## HIERARCHY OF SOLAR – LIKE SYSTEMS: EXTERNAL RADIUSES

| Layers of Matter | Levels of Matter $U^{(k)}$ | | OBSERVATIONS |
|---|---|---|---|
| | $a^{(k)}=\chi^k a_{ss,p}$; $\chi=3.66(6)$ | External Radiuses | (Ref.: LSS, Allen, Peebles, Rudnicki, Kalinkov, Holopov, Sharov, etc.) |
| E L A S | $a^{(32)}=\chi^{32} a_{ss,p}$ | 217.5·10⁶ Mpc | |
| | $a^{(31)}=\chi^{31} a_{ss,p}$ | 59.319·10⁶ Mpc | |
| | $a^{(30)}=\chi^{30} a_{ss,p}$ | 16.177·10⁶ Mpc | |
| | $a^{(29)}=\chi^{29} a_{ss,p}$ | 4.412·10⁶ Mpc | |
| | $a^{(28)}=\chi^{28} a_{ss,p}$ | 1203318 Mpc | |
| | $a^{(27)}=\chi^{27} a_{ss,p}$ | 328177 Mpc | |
| | $a^{(26)}=\chi^{26} a_{ss,p}$ | 89503 Mpc | |
| | $a^{(25)}=\chi^{25} a_{ss,p}$ | 24409.9 Mpc | |
| | $a^{(24)}=\chi^{24} a_{ss,p}$ | 6657.24 Mpc | |
| | $a^{(23)}=\chi^{23} a_{ss,p}$ | 1815.612 Mpc | |
| | $a^{(22)}=\chi^{22} a_{ss,p}$ | 495.167 Mpc | 421 Mpc [LSS:Rudnicki,Zieba] |
| | $a^{(21)}=\chi^{21} a_{ss,p}$ | 135.045 Mpc | 128 Mpc [LSS:Rudnicki,Zieba]; 120÷150 Mpc [LSS:Kalinkov et al.] |
| | $a^{(20)}=\chi^{20} a_{ss,p}$ | 36.830 Mpc | 35 Mpc [LSS:Rudnicki,Zieba]; 40÷50 Mpc [LSS:Kalinkov et al.] |
| G A L A X I E S | $a^{(19)}=\chi^{19} a_{ss,p}$ | 10.044 Mpc | $r_0$~5h⁻¹Mpc~7÷10 Mpc [LSS:Peebles]; 5.7 Mpc (Pis) [Allen] |
| | $a^{(18)}=\chi^{18} a_{ss,p}$ | 2.73946 Mpc | 2÷3.5 Mpc [LSS:Peebles]; |
| | $a^{(17)}=\chi^{17} a_{ss,p}$ | 0.7471 Mpc | 0.57 Mpc (Peg) [Allen] |
| | $a^{(16)}=\chi^{16} a_{ss,p}$ | 0.2037 Mpc | 0.15 Mpc (Her) [Allen] |
| | $a^{(15)}=\chi^{15} a_{ss,p}$ | 55.571 kpc | 46÷68 kpc (MW) [(Kukarkin; Holopov]; 30÷36 kpc (MW) [Zinn]; 35÷50 kpc (M31 And) [Sharov]; |
| | $a^{(14)}=\chi^{14} a_{ss,p}$ | 15.155 kpc | 11.5 kpc (M1O1) [Allen] |
| | $a^{(13)}=\chi^{13} a_{ss,p}$ | 4.133 kpc | 3.5 kpc (M82), 4 kpc (M104), 4.5 kpc (M51) [Allen] |
| | $a^{(12)}=\chi^{12} a_{ss,p}$ | 1.12729 kpc | 1 kpc (NGS 6822,205) [Allen] |
| | $a^{(11)}=\chi^{11} a_{ss,p}$ | 0.3074 kpc | 0.5 kpc (M32,NGC 147, 185) [Allen] |
| | $a^{(10)}=\chi^{10} a_{ss,p}$ | 83.848 pc | 80 pc ($\chi$,h Per); 64pc(M3)[Holopov]; 70pc(Gum); 85pc(NGC 2070) Diffuse Nebula; |
| S T A R S | $a^{(9)}=\chi^9 a_{ss,p}$ | 22.867 pc | 19pc(NGC 7243) [Holopov]; 16pc (NGC 2419)Globular Cluster [Allen, p.400]; 50pc Sco-Cen) –Open Cluster [Allen, p.396]; |
| | $a^{(8)}=\chi^8 a_{ss,p}$ | 6.236 pc | 6.9pc(M44) [Holopov]; 6pc(Per); 7pc($\chi$ Per) Open Cluster [Allen, p.396]; |
| | $a^{(7)}=\chi^7 a_{ss,p}$ | 1.7009 pc | 1.5pc (NGC 2632, I2602); 2pc (NGC2632,2682,4755,6531) Open clusters |
| | $a^{(6)}=\chi^6 a_{ss,p}$ | 0.4638 pc | 0.3pc (M78); 0.5pc (NGC 7023) – Diffuse Nebula; |
| | $a^{(5)}=\chi5^2 a_{ss,p}$ | 0.1265 pc | 0.1pc(NGC 3132,6720); 0.15pc (NGC6853) – Planetary Nebula; 0.15pc (NGC 2261) – Diffuse Nebula; |
| | $a^{(4)}=\chi^4 a_{ss,p}$ | 0.0345 pc | 0.03pc (NGC 7662); 0.035pc (NGC 7027); 0.04pc (NGC 3918, 6210, 6543, 7009); [Allen] |
| | $a^{(3)}=\chi^3 a_{ss,p}$ | 1940.9 AU | |
| | $a^{(2)}=\chi^2 a_{ss,p}$ | 529.35 AU | |
| | $a^{(1)}=\chi^1 a_{ss,p}$ | 144.37 AU | |
| | $a^{(0)}=\chi^0 a_{ss,p}$ | 39.37364 AU | Solar System: Radius of Pluto Orbit |